\documentclass[12pt]{article}
\usepackage{amsmath}
\usepackage{amsthm}
\usepackage{amssymb}
\usepackage[english]{babel}
\usepackage[dvips]{graphicx}

\makeatletter \@addtoreset{equation}{section}
\renewcommand{\theequation}{\thesection.\@arabic\c@equation}
\makeatother

\newtheorem{theorem}{Theorem}
\newtheorem{remark}{Remark}

\newcommand{\No}{N}

\begin{document}
\title{Separation of variables in the generalized \\ 4th Appelrot
class\thanks{Regular and Chaotic Dynamics, 2007, Vol. 12, No. 3, pp.
267-280}}

\author{Mikhail P. Kharlamov\\
\it\small Gagarin Street 8, Volgograd, 400131 Russia\\
\it\small e-mail: mharlamov@vags.ru}

\date{28.01.07}

\maketitle

\begin{abstract}
We consider the analogue of the 4th Appelrot class of motions of the
Kowalevski top for the case of two constant force fields. The
trajectories of this family fill the four-dimensional surface
$\mathfrak{O}$ in the six-dimensional phase space. The constants of
three first integrals in involution restricted to this surface fill
one of the sheets of the bifurcation diagram in ${\bf R}^3$. We
point out the pair of partial integrals to obtain the explicit
parametric equations of this sheet. The induced system on
$\mathfrak{O}$ is shown to be Hamiltonian with two degrees of
freedom having the thin set of points where the induced symplectic
structure degenerates. The region of existence of motions in terms
of the integral constants is found. We provide the separation of
variables on $\mathfrak{O}$ and the algebraic formulae for the
initial phase variables.
\end{abstract}

\footnotesize



\noindent DOI: 10.1134/S1560354707030021

\normalsize


\section {Preliminaries}
The equations of motion of the Kowalevski top in two constant fields, expressed in the
reference system $O{\bf e}_1{\bf e}_2{\bf e}_3$ of the principal axes of inertia
at the fixed point $O$,
\begin{equation}\label{s1n1}
\begin{array}{c}
2\dot\omega _1   = \omega _2 \omega _3  + \beta _3
,\;\quad 2\dot\omega _2 =  - \omega _1 \omega _3  -
\alpha _3 ,\; \quad
\dot\omega _3   = \alpha _2  - \beta _1 , \\
\dot\alpha _1   = \alpha _2 \omega _3  - \alpha _3
\omega _2 ,\; \quad \dot\beta _1   = \beta _2 \omega
_3  - \beta _3 \omega _2, \\
\dot\alpha _2   = \alpha _3 \omega _1  - \alpha _1
\omega _3 ,\; \quad \dot\beta _2   = \beta _3
\omega_1  -
\beta _1 \omega _3, \\
\dot\alpha_3   = \alpha_1 \omega_2  - \alpha_2
\omega_1 ,\; \quad \dot\beta_3   = \beta_1 \omega_2
- \beta_2 \omega_1
\end{array}
\end{equation}
can without loss of generality be restricted to the phase space $P^6
\subset {\bf{R}}^9 ({\boldsymbol {\omega }},{\boldsymbol {\alpha
}},{\boldsymbol {\beta }})$ defined by the geometric integrals
\begin{equation}\label{s1n2}
\alpha _1^2  + \alpha _2^2  + \alpha _3^2  = a^2 ,\quad \beta _1^2 +
\beta _2^2  + \beta _3^2  = b^2 ,\quad \alpha _1 \beta _1  + \alpha
_2 \beta _2  + \alpha _3 \beta _3  = 0
\end{equation}
(see \cite{KhRCD} for details). This system is completely integrable
due to the existence of three integrals in involution
\cite{Bogo,ReySem}
\begin{equation}\label{s1n3}
\begin{array}{l}
\displaystyle{H = \omega _1^2 + \omega _2^2 + \frac{1} {2}\omega
_3^2 - (\alpha _1
+ \beta _2 ),} \\[3mm]
K = (\omega _1^2 - \omega _2^2 + \alpha _1 - \beta _2 )^2 + (2\omega
_1 \omega _2 + \alpha _2 + \beta _1 )^2,\\[3mm]
\displaystyle{G = \frac{1}{4}(M_\alpha^2 + M_\beta^2 ) +
\frac{1}{2}\omega_3 M_\gamma - b^2 \alpha_1 - a^2 \beta_2,}
\end{array}
\end{equation}
where
\begin{equation}\label{s1n4}
\begin{array}{l}
M_\alpha = 2\omega_1 \alpha_1 + 2\omega_2 \alpha_2 +
\omega_3 \alpha_3 , \\
M_\beta = 2\omega_1 \beta_1 + 2\omega_2 \beta_2 + \omega_3 \beta_3 , \\
M_\gamma = 2\omega_1 (\alpha_2 \beta_3 - \alpha_3 \beta_2 ) +
2\omega_2 (\alpha_3 \beta_1 - \alpha_1 \beta_3 ) + \omega_3
(\alpha_1 \beta_2 - \alpha_2 \beta_1 ).
\end{array}
\end{equation}

For the general case
\begin{equation}\label{s1n5}
a > b > 0
\end{equation}
the explicit integration has not been found yet. It is natural to
study first the invariant submanifolds in $P^6$ such that the
induced system has only two degrees of freedom. It is proved in
\cite{KhMTT,KhRCD} that there exist only three submanifolds
$\mathfrak{M}, \mathfrak{N}, \mathfrak{O}$ of this type. The union
$\mathfrak{M} \cup \mathfrak{N} \cup \mathfrak{O}$ coincides with
the set of critical points of the integral map
\begin{equation}\label{s1n6}
H \times K \times G:P^6  \to {\bf{R}}^3.
\end{equation}

If $b=0$ (the classical Kowalevski case \cite{Kowa}), then the
critical set of the map (\ref{s1n6}) consists of the motions that
belong to the so-called four Appelrot classes \cite{Appel}. The set
$\mathfrak{M}$, first found in \cite{Bogo} as the zero level of the
integral $K$, generalizes the 1st Appelrot class. The phase topology
of the system induced on $\mathfrak{M}$ was studied in
\cite{ZotRCD}. The dynamical system on $\mathfrak{N}$ generalizing
the 2nd and 3rd Appelrot classes was explicitly integrated
in~\cite{KhSavUMB}. The present work considers the restriction of
the system~(\ref{s1n1}) to the invariant subset $\mathfrak{O}$.

Introduce the complex phase variables \cite{KhOne} ($i^2=-1$):
\begin{equation}\label{s1n7}
\begin{array}{l}
x_1 = (\alpha_1  - \beta_2) + i(\alpha_2  + \beta_1),\quad
x_2 = (\alpha_1  - \beta_2) - i(\alpha_2  + \beta_1 ), \\
y_1 = (\alpha_1  + \beta_2) + i(\alpha_2  - \beta_1), \quad y_2 =
(\alpha_1  + \beta_2) -
i(\alpha_2  - \beta_1), \\
 z_1 = \alpha_3  + i\beta_3, \quad
z_2 = \alpha_3  - i\beta_3,\\
w_1 = \omega_1  + i\omega_2 , \quad w_2 = \omega_1  - i\omega_2,
\quad w_3 = \omega_3.
\end{array}
\end{equation}
The system (\ref{s1n1}) takes the form
\begin{equation}\label{s1n8}
\begin{array}{c}
\begin{array}{ll}
{x'_1  =  - x_1 w_3  + z_1 w_1,} & {x'_2  = x_2 w_3  - z_2 w_2,} \cr
{y'_1  =  - y_1 w_3  + z_2 w_1,} & {y'_2  = y_2 w_3  - z_1 w_2 ,}
\cr {2z'_1  = x_1 w_2  - y_2 w_1,} & {2z'_2  =  - x_2 w_1 + y_1
w_2,}
\end{array} \\
2w'_1  =  - (w_1 w_3  + z_1 ),\quad 2w'_2  = w_2 w_3  + z_2, \quad
2w'_3 = y_2  - y_1.
\end{array}
\end{equation}
Here the prime stands for $d/d(it)$.

The set $\mathfrak{O}$, by the definition given in the work
\cite{KhRCD}, includes as a proper subset the following points
\begin{equation}\label{s1n9}
w_1 = w_2 = 0,\quad z_1 = z_2 = 0.
\end{equation}
The invariant relations (\ref{s1n9}) lead to the family of pendulum
motions first found in \cite{KhMTT}
\begin{equation}\label{s1n10}
\begin{array}{l}
{\boldsymbol {\alpha }} = a({\bf{e}}_1 \cos \theta  - {\bf{e}}_2
\sin \theta ),\quad {\boldsymbol {\beta }} = \pm \, b({\bf{e}}_1
\sin
\theta  + {\bf{e}}_2 \cos \theta ),\\[2mm]
{\boldsymbol\alpha}\times{\boldsymbol\beta} \equiv  \pm \, a
b\,{\bf{e}}_3 , \quad \displaystyle{{\boldsymbol {\omega }} =
\frac{d\theta}{dt}\, {\bf{e}}_3 ,\quad \frac{d^2\theta}{dt^2} =  -
(a \pm b)\sin \theta.}
\end{array}
\end{equation}
The corresponding values of the integrals (\ref{s1n3}) satisfy one
of the following
\begin{equation}\label{s1n11}
g = \pm \, ab\,h,\quad k = (a \mp b)^2 ,\quad h \geqslant  - (a \pm
b).
\end{equation}

In the sequel we use functions and expressions having singularities
at the points (\ref{s1n9}). Therefore, by default, we exclude the
trajectories (\ref{s1n10}). The remaining part of $\mathfrak{O}$ can be
described by the pair of equations
\begin{equation}\label{s1n12}
R_1 = 0, \quad R_2=0,
\end{equation}
where
\begin{equation}\label{s1n13}
\begin{split}
R_1 & =\displaystyle{\frac{w_2 x_1+w_1 y_2+w_3 z_1}{w_1}-\frac{w_1
x_2+w_2 y_1+w_3
z_2}{w_2},} \\
R_2 & = \displaystyle{(w_2 z_1+w_1 z_2)w_3^2+\Bigl[\frac{w_2
z_1^2}{w_1}+\frac{w_1
z_2^2}{w_2}+w_1 w_2(y_1+y_2)+x_1 w_2^2+x_2 w_1^2\Bigr]w_3 +}\\
    & \phantom{=} + \displaystyle{\frac{w_2^2 x_1 z_1}{w_1} + \frac{w_1^2 x_2
z_2}{w_2}+ x_1 z_2 w_2+ x_2 z_1 w_1 +(w_1 z_2-w_2 z_1)(y_1-y_2).}
\end{split}
\end{equation}
For the derivatives we have, in virtue of (\ref{s1n8}),
\begin{equation}\label{s1n14}
\begin{array}{c}
R_1'=\kappa_2 R_2, \qquad R_2'= \kappa_1 R_1, \\[3mm]
\displaystyle{\kappa_1 = \frac{1}{2w_1 w_2}\bigl[(w_1 w_2 w_3 + z_2
w_1+z_1 w_2)^2 + w_1 w_2(x_2 w_1^2+x_1 w_2^2)\bigr], \quad
\kappa_2=\frac{1}{2w_1 w_2}.}
\end{array}
\end{equation}

The equations (\ref{s1n14}) straightforwardly prove that the set
(\ref{s1n12}) is preserved by the phase flow (\ref{s1n1}). Besides,
it follows that the Poisson bracket $\{R_1,R_2\}$ is a partial
integral on $\mathfrak{O}$. The subset in $\mathfrak{O}$ defined by
the equation
\begin{equation}\label{s1n15}
\{R_1,R_2\}=0
\end{equation}
is the set of points at which the induced symplectic structure is
degenerate. Below we calculate $\{R_1,R_2\}$ using the appropriate
integrals on $\mathfrak{O}$ and show that the set (\ref{s1n15}) is
of measure zero. Thus, the dynamical system induced on
$\mathfrak{O}$ by the system (\ref{s1n1}) is almost everywhere a
Hamiltonian system with two degrees of freedom. In particular,
almost all its integral manifolds consist of two-dimensional
Liouville tori.

\section{Partial integrals}
Recall that in the classical Kowalevski case (${\boldsymbol {\beta
}} = 0$) there exists the momentum integral
\begin{equation}\label{s2n1}
L = \frac {1}{2} {\bf I}{\boldsymbol\omega}{\boldsymbol\cdot}
{\boldsymbol\alpha} \qquad ({\bf I}=\mathrm{diag}\,\{2,2,1\}).
\end{equation}
Then the integral $G$ becomes equal to $L^2 $. Let $\ell $ denote
the constant of the integral (\ref{s2n1}). In Appelrot's notation
the 4th class of {\it especially remarkable motions} is defined by
the following conditions.

(i) The second polynomial of Kowalevski has a multiple root. One of
the Kowalevski variables remains constant and equal to the multiple
root $s$ of the corresponding Euler resolvent $\varphi (s) = s(s -
h)^2  + (a^2 - k)s - 2\ell ^2$:
\begin{equation}\label{s2n2}
\varphi (s) = 0,\quad \varphi '(s) = 0.
\end{equation}

(ii) Two equatorial components of the angular velocity are constant:
${\omega_1 \equiv -\ell/s}$, ${\omega _2 \equiv 0}$. Given
${{\boldsymbol \beta}=0}$, this fact can be written in the form
\begin{equation}\label{s2n3}
\frac {{\bf
I}{\boldsymbol\omega}{\boldsymbol\cdot}{\boldsymbol\alpha}} { {\bf
I}{\boldsymbol\omega}{\boldsymbol\cdot}{\bf e}_1 } = -s, \quad {\bf
I}{\boldsymbol\omega}{\boldsymbol\cdot}{\boldsymbol\beta}=0, \quad
{\bf I}{\boldsymbol\omega}{\boldsymbol\cdot}{\bf e}_2 =0.
\end{equation}

The next statement establishes the conditions similar to
(\ref{s2n3}) for the generalized top.

\begin{theorem} On each trajectory belonging to
$\mathfrak{O}$ the ratios
$$
\frac {{\bf
I}{\boldsymbol\omega}{\boldsymbol\cdot}{\boldsymbol\alpha}} { {\bf
I}{\boldsymbol\omega}{\boldsymbol\cdot}{\bf e}_1 }, \quad \frac {
{\bf I}{\boldsymbol\omega}{\boldsymbol\cdot}{\boldsymbol\beta}} {
{\bf I}{\boldsymbol\omega}{\boldsymbol\cdot}{\bf e}_2 }
$$
are constant and equal to each other.
\end{theorem}
\begin{proof} Let
\begin{equation}\label{s2n4}
{\bf{M}} = {\bf{I}\boldsymbol {\omega}}, \quad M_j  =
{\bf{I}\boldsymbol {\omega}}{\boldsymbol\cdot}{\bf{e}}_j \quad
(j=1,2,3).
\end{equation}
Due to (\ref{s1n4}) we also have $ M_\alpha   = {\bf{I}\boldsymbol
{\omega}}{\boldsymbol\cdot}{\boldsymbol {\alpha }}$, $M_\beta =
{\bf{I}\boldsymbol {\omega}}{\boldsymbol\cdot}{\boldsymbol {\beta
}}$. Then the first equation (\ref{s1n12}) yields
\begin{equation}\label{s2n5}
\frac{M_\alpha}{M_1} - \frac{M_\beta}{M_2} = 0.
\end{equation}
Introduce the function
$$
S =  - \frac{M_\alpha  M_1  + M_\beta  M_2 } {M_1^2  + M_2^2 }
$$
and calculate its time derivative:
$$
\frac{dS}{dt} = - \frac{(M_1^2  + M_2^2) \omega _3  + 4\alpha _3 M_1
+ 4\beta _3 M_2 }{2(M_1^2  + M_2^2 )^2}(M_\alpha  M_2 - M_\beta
M_1).
$$
In virtue of (\ref{s2n5}) the right hand side is identically zero.
Therefore $S$ is a partial integral on $\mathfrak{O}$. Denote
by~$s$ the corresponding constant:
\begin{equation}\label{s2n6}
\frac{M_\alpha  M_1  + M_\beta  M_2 } {M_1^2  + M_2^2 } =  - s.
\end{equation}
From (\ref{s2n5}), (\ref{s2n6}) we obtain
\begin{equation}\label{s2n7}
M_\alpha = - s M_1, \quad M_\beta =  - s M_2
\end{equation}
with the constant value $s$.
\end{proof}

Note that according to {\rm{(\ref{s2n7})}} the function $S$ can be
written in either of the representations
\begin{equation}\label{s2n8}
\displaystyle{S =  - \frac{1}{4}\left(\frac{M_ \alpha+ i
M_\beta}{\omega_1+i \omega_2} + \frac{M_ \alpha - i
M_\beta}{\omega_1 - i \omega_2}\right) = - \frac{1}{2}\frac{M_
\alpha+ i M_\beta}{\omega_1+i \omega_2} = - \frac{1}{2}\frac{M_
\alpha- i M_\beta}{\omega_1-i \omega_2}.}
\end{equation}

\begin{theorem} On the set $\mathfrak{O}$ the system
{\rm{(\ref{s1n1})}} has the partial integral
\begin{equation}\label{s2n9}
\displaystyle{{\rm T} = \frac {1}{2} (M_\alpha M_1 + M_\beta  M _2
)- 2(\alpha _1 \beta _2  - \alpha _2 \beta _1 ) + a^2  + b^2 .}
\end{equation}
\end{theorem}
\begin{proof} The time derivative of $\mathrm{T}$
$$
\frac {d{\rm T}}{dt} = \frac{1}{4}\,\omega _3 (M_\alpha  M_2 -
M_\beta M_1)
$$
vanishes on $\mathfrak{O}$ due to (\ref{s2n5}).
\end{proof}

Denote by $\tau $ the constant of the integral ${\rm T}$.

\begin{remark}\label{rem2}
In the work {\rm{\cite{KhRCD}}} equations similar to
{\rm{(\ref{s1n12})}} were derived from the condition that the
function
\begin{equation}\label{s2n10}
2G + (\tau  - a^2-b^2 )H + s\,K
\end{equation}
with Lagrange's multipliers $s,\tau $ has a critical point on $P^6$.
Using the coordinates {\rm{(\ref{s1n7})}} we have from
{\rm{(\ref{s2n8})}}, {\rm{(\ref{s2n9})}}
\begin{eqnarray}
& & \begin{array}{l} \displaystyle{S=-\frac{w_1(x_2 w_1+y_1 w_2+z_2
w_3)+w_2(y_2 w_1+x_1
w_2+z_1 w_3)}{4 w_1 w_2}} = \\[4mm]
\quad  = \displaystyle{-\frac{x_2 w_1+y_1 w_2+z_2 w_3}{2 w_2}
= - \frac {y_2 w_1+x_1 w_2+z_1 w_3} {2 w_1},}
\end{array} \label{s2n11} \\
& & \begin{array}{l} \displaystyle{{\rm T}=\frac{1}{2}[w_1(x_2 w_1+y_1
w_2+z_2 w_3)+w_2(y_2
w_1+x_1 w_2+z_1 w_3)]+x_1 x_2+z_1 z_2 = }\\[4mm]
\quad = -2 S\, w_1 w_2 +x_1 x_2+z_1 z_2.
\end{array}\label{s2n12}
\end{eqnarray}
Comparing {\rm{(\ref{s2n11})}}, {\rm{(\ref{s2n12})}} with the
expressions for $s,\tau$ given in {\rm{\cite{KhRCD}}} we see that on
$\mathfrak{O}$ Lagrange's multipliers in {\rm{(\ref{s2n10})}}
coincide with the constants of the integrals $S,{\rm T}$ introduced here.
\end{remark}

Supposing (\ref{s1n5}), introduce parameters $p,r$ ($p>r>0$)
such that
\begin{equation}\label{s2n13}
p^2  = a^2  + b^2 ,\quad r^2  = a^2  - b^2.
\end{equation}
Let $h,k,g$ denote the constants of the general integrals
(\ref{s1n3}). Then according to Remark~\ref{rem2}, from the results
of the work~\cite{KhRCD} we obtain that on $\mathfrak{O}$
\begin{equation}\label{s2n14}
h = \frac {p^2 -\tau}{2s}+ s, \quad k = \frac{\tau^2 - 2p^2\tau +
r^4}{4s^2}+\tau, \quad g = \frac{p^4-r^4} {4s}+ \frac{1}{2}(p^2
-\tau)s.
\end{equation}
These equations can be considered as parametric equations of the
corresponding bifurcation sheet of the integral map (\ref{s1n6}).
Eliminating $\tau$ we have
\begin{equation}\label{s2n15}
\psi (s) = 0,\quad \psi '(s) = 0,
\end{equation}
where
$$
\psi (s) = s^2 (s - h)^2  + (p^2  - k)s^2  - 2gs + \frac {p^4  -
r^4}{4}.
$$
If ${\boldsymbol {\beta }} = 0$ ($p^2  = r^2  = a^2 $), then $\psi
(s) = s\,\varphi (s)$ and the conditions (\ref{s2n15}) turn to
(\ref{s2n2}). Therefore, the set of trajectories belonging to
$\mathfrak{O}$ is the generalization of the set of the {\it
especially remarkable motions} of the 4th~Appelrot class.

\section{Parametric equations of integral manifolds}
Due to the equations (\ref{s2n14}) the functions $S,{\rm T}$ form
a complete system of first integrals on $\mathfrak{O}$. In
particular, the equations of the integral manifold
$$
\{ \zeta  \in P^6 : {H(\zeta ) = h},
{K(\zeta ) = k}, {G(\zeta ) = g}\}
$$
in this class of motions are equivalent to the relations
(\ref{s1n12}) and the equations
\begin{equation}\label{s4n1}
S = s,\; {\rm T} = \tau.
\end{equation}

Using (\ref{s2n5}), (\ref{s2n11}) we replace the equations $R_1=0$,
$S=s$ by
\begin{equation}\label{s4n2}
\begin{array}{l}
(y_2  + 2s)w_1  + x_1 w_2  + z_1 w_3  = 0,  \\
x_2 w_1  + (y_1  + 2s)w_2  + z_2 w_3  = 0.
\end{array}
\end{equation}
At the same time, from (\ref{s1n13}) and (\ref{s2n12}) the system
$R_2=0$, $\mathrm{T}=\tau$ is equivalent to
\begin{eqnarray}
& & x_2 z_1 w_1  + x_1 z_2 w_2  + (\tau  - x_1 x_2 )w_3  = 0,  \label{s4n3}\\
& & 2s \, w_1 w_2  - (x_1 x_2  + z_1 z_2 ) + \tau  = 0. \label{s4n4}
\end{eqnarray}
To obtain a closed system of equations we must add the geometric
integrals (\ref{s1n2}). In terms of the variables (\ref{s1n7}) we
get
\begin{eqnarray}
& & z_1^2  + x_1 y_2  = r^2, \qquad
z_2^2  + x_2 y_1  = r^2, \label{s4n5}\\
& &  x_1 x_2 + y_1 y_2  + 2z_1 z_2  = 2p^2\label{s4n6}.
\end{eqnarray}
The variables (\ref{s1n7}) by definition satisfy
\begin{equation}\label{s4n7}
x_2  = \overline{\mathstrut x_1}, \quad y_2  = \overline{\mathstrut
y_1},\quad z_2  = \overline{\mathstrut z_1}, \quad  w_2  =
\overline{\mathstrut w_1}, \quad  w_3 \in {\bf R},
\end{equation}
thus forming a space of real dimension 9. Seven relations
(\ref{s4n2}) -- (\ref{s4n6}) define the integral manifold. In the
case of independency of the integrals $S,{\rm T}$ this manifold is
two-dimensional and consists of Liouville tori filled with
quasi-periodic motions.

Let
\begin{equation}\label{s4n8}
x  = \sqrt{\mathstrut x_1 x_2} ,\quad z  = \sqrt{\mathstrut z_1
z_2}, \quad \xi = 2s \, w_1 w_2.
\end{equation}
Then from (\ref{s4n4})
\begin{equation}\label{s4n9}
\xi = x^2+z^2-\tau.
\end{equation}
From (\ref{s4n5}), (\ref{s4n8}) we obtain
\begin{equation}\label{s4n10}
(r^2  - x_1 y_2 )(r^2  - x_2 y_1 ) = z^4,
\end{equation}
and the equation (\ref{s4n6}) yields
\begin{equation}\label{s4n11}
y_1 y_2  = 2p^2  - x^2  - 2z^2.
\end{equation}
Hence
\begin{equation}\label{s4n12}
r^2 (x_1 y_2  + x_2 y_1 ) = r^4  + 2p^2 x^2  - (x^2  + z^2 )^2 .
\end{equation}
Rewrite (\ref{s4n5}) in the form
\begin{equation}\label{s4n13}
\begin{array}{l}
(z_1  + z_2 )^2  = 2r^2  - (x_1 y_2  + x_2 y_1 ) + 2z^2 ,  \\
(z_1  - z_2 )^2  = 2r^2  - (x_1 y_2  + x_2 y_1 ) - 2z^2
\end{array}
\end{equation}
and substitute (\ref{s4n12}) to obtain
\begin{equation}\label{s4n14}
r^2 (z_1  + z_2 )^2  = \Phi_ 1,\quad r^2 (z_1  - z_2 )^2  = \Phi_2,
\end{equation}
where
\begin{equation}\label{s4n15}
\begin{array}{l}
\Phi_1 = (x^2  + z^2  + r^2 )^2  - 2(p^2  + r^2 )x^2 = (\xi + \tau + r^2 )^2 - 2(p^2 + r^2)x^2, \\
\Phi_2 = (x^2  + z^2  - r^2 )^2  - 2(p^2  - r^2 )x^2 = (\xi + \tau -
r^2 )^2 - 2(p^2 - r^2)x^2.
\end{array}
\end{equation}

Note that the equilibria ${\boldsymbol \omega} \equiv 0$ of the
system (\ref{s1n1}) are included in the family of motions
(\ref{s1n10}). On the rest of the trajectories in $\mathfrak{O}$ the
determinant of the equations (\ref{s4n2}), (\ref{s4n3}) in $w_j$
($j=1,2,3$) vanishes identically. Calculate this determinant and
eliminate $z_1^2 ,z_2^2$ from (\ref{s4n5}) and the product $y_1 y_2$
from (\ref{s4n11}) to obtain
\begin{equation}\label{s4n16}
\begin{array}{l}
2s[(r^2 x_1  - \tau y_1 ) + (r^2 x_2  - \tau y_2 )] = - r^2 (x_1 y_2  + x_2 y_1 ) +\\
\quad  + 2[2s^2 (\tau  - x^2 ) + p^2 (\tau  + x^2 ) - \tau (x^2  +
z^2 )].
\end{array}
\end{equation}
On the other hand, the direct calculation in view of (\ref{s4n8}),
(\ref{s4n11}) gives
\begin{equation}\label{s4n17}
(r^2 x_1  - \tau y_1 )(r^2 x_2  - \tau y_2 ) = r^4 x^2  + \tau (2p^2
- x^2  - 2z^2 ) - r^2 \tau (x_1 y_2  + x_2 y_1 ).
\end{equation}
Denote
\begin{equation}\label{s4n18}
\sigma  = \tau ^2  - 2p^2 \tau  + r^4 ,\quad \chi  = \sqrt k
\geqslant 0.
\end{equation}
From the second relation (\ref{s2n14}) we have the identity
\begin{equation}\label{s4n19}
4s^2 \chi ^2  = \sigma  + 4s^2 \tau.
\end{equation}
Introduce the complex conjugate variables
\begin{equation}\label{s4n20}
\mu _1  = r^2 x_1  - \tau y_1 ,\quad \mu _2  = r^2 x_2  - \tau y_2.
\end{equation}
Eliminating $x_1 y_2  + x_2 y_1 $ from (\ref{s4n16}), (\ref{s4n17})
with the help of (\ref{s4n12}) we come to the system
\begin{equation}\label{s4n21}
\begin{array}{c}
2s(\mu _1  + \mu _2 ) = \xi^2  - 4s^2 (x^2 -\tau ) - \sigma, \\
\mu _1 \mu _2  = \tau \xi^2  + \sigma x^2  - \tau \sigma .
\end{array}
\end{equation}

Choose
\begin{equation}\label{s4n22}
\mu _1^*  = \sqrt {2s\mu _1  - 4s^2\tau } ,\quad \mu _2^*  = \sqrt
{2s\mu _2  - 4s^2\tau}
\end{equation}
to be complex conjugate. Then the system (\ref{s4n21}) takes the
form
\begin{equation}\label{s4n23}
(\mu _1^*  +\mu _2^* )^2  = \Psi_1,\quad (\mu _1^* - \mu _2^* )^2 =
\Psi_2,
\end{equation}
where
\begin{equation}\label{s4n24}
\begin{array}{l}
\Psi_1 = \xi^2  -4s^2(x + \chi )^2, \quad \Psi_2 = \xi^2  -4s^2(x -
\chi )^2.
\end{array}
\end{equation}

As was to be expected from dimensional reasoning, all phase variables could be expressed
in terms of two almost everywhere independent auxiliary variables.
The above formulae emphasize the special role of the pair $(x,\xi)$.

From (\ref{s4n22}), (\ref{s4n23}) we find
\begin{equation}\label{s4n25}
\displaystyle{\mu_1=2s\tau+\frac{1}{8s}(\sqrt{\Psi_1}+\sqrt{\Psi_2})^2,
\quad \mu_2=2s\tau+\frac{1}{8s}(\sqrt{\Psi_1}-\sqrt{\Psi_2})^2.}
\end{equation}
Further on the calculation sequence is as follows. From
(\ref{s4n14}) we find $z_1,z_2$. Multiplying the equations
(\ref{s4n20}) by $x_2,x_1$ respectively and using (\ref{s4n5}) we
get
$$
\begin{array}{l}
x_2 \mu_1 = r^2 x^2 -\tau x_2 y_1 = r^2(x^2 -\tau)+ \tau z_2^2, \\
x_1 \mu_2 = r^2 x^2 -\tau x_1 y_2 = r^2(x^2 -\tau)+ \tau z_1^2,
\end{array}
$$
whence $x_1, x_2$ are found. Substituting these values back to
(\ref{s4n20}) we find $y_1, y_2$. As a result, after some obvious
transformations we obtain the following expressions for the
configuration variables:
\begin{eqnarray}
& &
\begin{array}{l}
\displaystyle{x_1=\frac{2s}{r^2}
\frac{4r^4(x^2-\tau)+\tau(\sqrt{\mathstrut \Phi_1}+\sqrt{\mathstrut
\Phi_2})^2}
{16s^2\tau+(\sqrt{\mathstrut \Psi_1}-\sqrt{\mathstrut \Psi_2})^2},} \\[4mm]
\displaystyle{x_2=\frac{2s}{r^2}
\frac{4r^4(x^2-\tau)+\tau(\sqrt{\mathstrut \Phi_1}-\sqrt{\mathstrut
\Phi_2})^2} {16s^2\tau+(\sqrt{\mathstrut \Psi_1}+\sqrt{\mathstrut
\Psi_2})^2},}
\end{array} \label{s4n26} \\
& & \begin{array}{l} \displaystyle{y_1= 2s \frac{4[2 \tau \xi - \tau
(x^2-\tau)+ \sigma] - (\sqrt{\mathstrut \Phi_1}-\sqrt{\mathstrut
\Phi_2})^2} {16s^2\tau+(\sqrt{\mathstrut \Psi_1}-\sqrt{\mathstrut
\Psi_2})^2},}
\\[4mm]
\displaystyle{y_2=2s \frac{4[2 \tau \xi - \tau (x^2-\tau)+ \sigma] -
(\sqrt{\mathstrut \Phi_1}+\sqrt{\mathstrut \Phi_2})^2}
{16s^2\tau+(\sqrt{\mathstrut \Psi_1}+\sqrt{\mathstrut \Psi_2})^2},}
\end{array} \label{s4n27}\\
& & \, \, \displaystyle{z_1=\frac{1}{2r} (\sqrt{\mathstrut \Phi_1} +
\sqrt{\mathstrut \Phi_2}), \quad z_2=\frac{1}{2r} ( \sqrt{\mathstrut
\Phi_1} - \sqrt{\mathstrut \Phi_2})}.\label{s4n28}
\end{eqnarray}

Note that all radicals are algebraic. The formal choice of signs in
(\ref{s4n26})--(\ref{s4n28}) is defined by the initial choice in the
expressions for $\mu_1, \mu_2, z_1, z_2$. The polynomials $\Phi_{j},
\Psi_{j}$ $(j=1,2)$ obviously split to multipliers linear with
respect to $x,\xi$. Then, typically, the projection of an integral
manifold onto the $(x,\xi)$-plane has the form of a quadrangle. Fix
any inner point $(x,\xi)$ of such projection. Then the expressions
(\ref{s4n26})--(\ref{s4n28}) define eight points of the
configuration space with different set of signs of the radicals
$\sqrt{\mathstrut \Phi_1}, \sqrt{\mathstrut \Phi_2},
\sqrt{\mathstrut \Psi_1 \Psi_2}$.

To find $w_3$, use the energy integral. On account of (\ref{s2n14})
it takes the form
$$
2 s\, w_3^2+4s\, w_1 w_2-2s(y_1+y_2)=4s^2+2p^2-2\tau.
$$
Substitute $2 s\, w_1 w_2$ from (\ref{s4n4}) and replace $2p^2$ by
its expression from (\ref{s4n6}) to obtain
\begin{equation}\label{s4n29}
2 s\, w_3^2=D,
\end{equation}
where
\begin{equation}\label{s4n30}
D = (y_1+2s)(y_2+2s)-x_1 x_2
\end{equation}
is the determinant of (\ref{s4n2}) with respect to $w_1,w_2$. In
particular, this determinant vanishes along with $w_3$. Due to this
fact we can express $w_1, w_2$ either as linear functions of $w_3$,
or in inverse proportion to $w_3$:
\begin{equation}\label{s4n31}
\begin{array}{l}
\displaystyle{w_1=\frac{x_1
z_2-(y_1+2s)z_1}{(y_1+2s)(y_2+2s)-x^2}w_3=\frac{x_1
z_2-(y_1+2s)z_1}{2s\, w_3},}\\
\displaystyle{w_2=\frac{x_2
z_1-(y_2+2s)z_2}{(y_1+2s)(y_2+2s)-x^2}w_3=\frac{x_2
z_1-(y_2+2s)z_2}{2s\, w_3}.}
\end{array}
\end{equation}

With two possibilities of the sign choice for $w_3$ in (\ref{s4n29})
we have that the inverse image in the integral manifold
${\{S=s,\mathrm{T}=\tau\}\cap \mathfrak{O}}$ of a generic point
$(x,\xi)$ contains 16 points.

We also need the explicit formula for $w_3$ in terms of $x,\xi$.
From (\ref{s4n27}) we write
\begin{equation}\label{s4n32}
\begin{array}{l} \displaystyle{y_1+2s = 2s \frac{4[2 \tau \xi - \tau
(x^2-\tau)+ 4 s^2\chi^2] - (\sqrt{\mathstrut
\Phi_1}-\sqrt{\mathstrut \Phi_2})^2 +(\sqrt{\mathstrut
\Psi_1}-\sqrt{\mathstrut \Psi_2})^2} {16s^2\tau+(\sqrt{\mathstrut
\Psi_1}-\sqrt{\mathstrut \Psi_2})^2},}
\\[4mm]
\displaystyle{y_2+2s=2s \frac{4[2 \tau \xi - \tau (x^2-\tau)+ 4
s^2\chi^2] - (\sqrt{\mathstrut \Phi_1}+\sqrt{\mathstrut
\Phi_2})^2+(\sqrt{\mathstrut \Psi_1}+\sqrt{\mathstrut \Psi_2})^2}
{16s^2\tau+(\sqrt{\mathstrut \Psi_1}+\sqrt{\mathstrut \Psi_2})^2}.}
\end{array}
\end{equation}
Note that
\begin{equation}\label{s4n33}
[{16s^2\tau+(\sqrt{\mathstrut \Psi_1}-\sqrt{\mathstrut
\Psi_2})^2}][{16s^2\tau+(\sqrt{\mathstrut \Psi_1}+\sqrt{\mathstrut
\Psi_2})^2}]=64s^2(\tau \xi^2+\sigma x^2-\tau\sigma).
\end{equation}
Then
$$
D=\frac{\sqrt{\mathstrut \Phi_1 \Phi_2 \Psi_1 \Psi_2}-P}{2 (\tau
\xi^2+\sigma x^2-\tau\sigma)},
$$
where
$$
\begin{array}{rcl}
P(x,\xi)&=&\xi^4+2\tau\xi^3+2[(\tau-2s^2-p^2)x^2-\tau(2s^2-p^2)-r^4]\xi^2-\\
& & -8s^2[(\tau-2\chi^2)x^2+\tau\chi^2]\xi -4s^2
(x^2-\chi^2)[2(\tau-p^2)x^2-(\tau^2-r^4)].
\end{array}
$$
Let
$$
\begin{array}{l}
Q(x,\xi)=(\,\xi+ \tau +2s^2 - p^2)^2-4s^2x^2 +r^4 -(2s^2 - p^2)^2, \\
P_1(x,\xi)= P(x,\xi)+2x Q(x,\xi) \sqrt{\tau \xi^2+\sigma x^2-\tau\sigma},\\
P_2(x,\xi)= P(x,\xi)-2x Q(x,\xi) \sqrt{\tau \xi^2+\sigma
x^2-\tau\sigma}.
\end{array}
$$
Direct calculation proves the identity $ P_1 P_2\equiv \Phi_1 \Phi_2
\Psi_1 \Psi_2$. On the other hand, $P_1+P_2\equiv 2P$. Therefore,
$$
D=-\frac{({\sqrt{\mathstrut P_1}-\sqrt{\mathstrut P_2}})^2}{4 (\tau
\xi^2+\sigma x^2-\tau\sigma)},
$$
whence
\begin{equation}\label{s4n34}
w_3=i\frac{{\sqrt{\mathstrut P_1}-\sqrt{\mathstrut
P_2}}}{2\sqrt{2s\,(\tau \xi^2+\sigma x^2-\tau\sigma)}}.
\end{equation}

Below we use one more representation of $w_1,w_2$ not depending on
$w_3$. The system (\ref{s4n2}) yields
$$
x_2 w_1^2=-z_2 w_1 w_3-(y_2+2s) w_1 w_2,\quad x_1 w_2^2=-z_1 w_2
w_3-(y_1+2s) w_1 w_2.
$$
Substitute the expressions for $w_1 w_3$, $w_2 w_3$ found from the
inverse proportion in (\ref{s4n31}) and eliminate $w_1 w_2$ by
(\ref{s4n4}) to obtain
$$
2s \, x_2 w_1^2 = - (\mu_1-2s \tau)-2s x^2,\quad 2s \, x_1 w_2^2 = -
(\mu_2-2s \tau)-2s x^2.
$$
Then from (\ref{s4n25}) we find
\begin{equation}\label{s4n35}
\displaystyle{w_1 = \frac{i}{4s\sqrt{x_2}}(\sqrt{\mathstrut
\Theta_1}+\sqrt{\mathstrut \Theta_2}), \quad w_2 =
\frac{i}{4s\sqrt{x_1}}(\sqrt{\mathstrut \Theta_1}-\sqrt{\mathstrut
\Theta_2}),}
\end{equation}
where
\begin{equation}\label{s4n36}
\Theta_1(x,\xi) =(\xi-2s x)^2-4 s^2 \chi ^2,\quad \Theta_2(x,\xi)
=(\xi + 2s x)^2-4 s^2 \chi ^2.
\end{equation}

Note that the products $\Theta_1 \Theta_2$ and $\Psi_1 \Psi_2$
coincide. The choice of the signs in (\ref{s4n35}) is determined by
the condition ${(\sqrt{x_2}w_1)(\sqrt{x_1}w_2)=x \,\xi /2s}$
following from (\ref{s4n8}). The signs of the complex conjugate
values $\sqrt{x_1}$, $\sqrt{x_2}$ must be chosen in such a way that
the expressions (\ref{s4n35}), (\ref{s4n35}) satisfy one of the
equations (\ref{s4n2}), (\ref{s4n3}) (then the other two hold
automatically).

Thus, all phase variables are algebraically expressed in terms of
two auxiliary variables $x, \xi$; the domain of the latter depends
on the constants of the first integrals.

\section{Singularities of the induced symplectic structure}
The Hamiltonian structure of the system (\ref{s1n1}) is provided by
the Poisson brackets on ${\bf R}^9(\boldsymbol \omega,\boldsymbol
\alpha,\boldsymbol \beta)$. In notation (\ref{s2n4}) these brackets
are \cite{Bogo}
\begin{equation}\label{s7n1}
\begin{array}{c}
\{M_j,M_k\}=\varepsilon_{j k l} M_l,\quad
\{M_j,\alpha_k\}=\varepsilon_{j k l} \alpha_l,\quad
\{M_j,\beta_k\}=\varepsilon_{j k l} \beta_l, \\
\{\alpha_j,\alpha_k\}=\{\alpha_j,\beta_k\}=\{\beta_j,\beta_k\}=0.
\end{array}
\end{equation}
Being restricted to $P^6$ they correspond to Kirillov's symplectic
form $\boldsymbol \lambda \in \Lambda^2(P^6)$. Recall the following
well-known facts. Let $N \subset P^6$ be a submanifold defined in
$P^6$ by two independent equations
\begin{equation}\label{s7n2}
f_1=0, \quad f_2=0
\end{equation}
and let $\mathfrak{X}_1, \mathfrak{X}_2$ be the Hamiltonian vector
fields with the Hamilton functions $f_1, f_2$. Then the span of
$\mathfrak{X}_1, \mathfrak{X}_2$ at each point $\zeta \in N$ is skew
orthogonal to the tangent space $T_\zeta N$. Therefore the
restriction of $\boldsymbol \lambda$ to $N$ is non-degenerate at the
point $\zeta$ if and only if
\begin{equation}\label{s7n3}
\{f_1,f_2\}(\zeta)=\boldsymbol \lambda (\mathfrak{X}_1,
\mathfrak{X}_2)(\zeta) \ne 0.
\end{equation}

Let us calculate the bracket $\{R_1, R_2\}$ of the functions
(\ref{s1n13}). As $R_1$ is pure imaginary denote
\begin{equation}\label{s7n4}
\displaystyle{R=\frac{1}{i}\{R_1, R_2\}.}
\end{equation}

The change of variables (\ref{s1n7}) is linear with constant
coefficients, so the rules (\ref{s7n1}) are easily transformed for
new coordinates. Omitting technical details we present $R$ in the
form
\begin{equation}\label{s7n5}
R=\frac{F_1}{w_1^3w_2^3w_3^2}[w_3(w_3 F_2+F_3)^2-F_4],
\end{equation}
where
\begin{equation}\label{s7n6}
\begin{array}{l}
F_1=z_1 z_2 w_3+x_2 z_1 w_1+x_1 z_2 w_2,\\[2mm]
F_2=w_1 w_2 w_3+ z_2 w_1+ z_1 w_2,\\[2mm]
F_3=2 w_1^2 w_2^2+x_2 w_1^2+x_1 w_2^2,\\[2mm]
F_4=4 w_1^2 w_2^2 [(w_1^2 w_2^2+x_2 w_1^2+x_1 w_2^2)w_3+x_2 z_1
w_1+x_1 z_2 w_2].
\end{array}
\end{equation}
The variables $y_1, y_2$ have been eliminated as the solution of the
system (\ref{s1n12})
\begin{equation}\label{s7n7}
\displaystyle{y_1=-\frac{w_1 w_3(x_2 w_1+z_2 w_3)+F_1}{w_1 w_2
w_3},}\quad \displaystyle{y_2=-\frac{w_2 w_3(x_1 w_2+z_1
w_3)+F_1}{w_1 w_2 w_3}.}
\end{equation}

Fixing the values $h,k,g,s,\tau$ of the functions (\ref{s1n3}),
(\ref{s2n11}), (\ref{s2n12}) we have
\begin{equation}\label{s7n8}
w_3 F_2+F_3=2 w_1 w_2 h-2(x_1 x_2+z_1 z_2 - \tau), \quad F_4=4w_1^2
w_2^2 w_3 (\tau-k).
\end{equation}
Then from (\ref{s2n14}), (\ref{s4n3}), (\ref{s4n4}), (\ref{s4n8}),
(\ref{s4n9}) we obtain
\begin{equation}\label{s7n9}
\begin{array}{l}
\displaystyle{w_3 F_2+F_3=\frac{\xi}{s}(\frac{p^2-\tau}{2s}-s),
\quad F_4=-\frac{\sigma \xi^2 w_3}{4s^4},} \quad \displaystyle{F_1 =
\xi w_3, \quad w_1^3 w_2^3 =\frac{\xi^3}{8s^3}.}
\end{array}
\end{equation}
Finally, the expression (\ref{s7n5}) takes the form
\begin{equation}\label{s7n10}
\displaystyle{R=\frac{8}{s}\left[s^4-(p^2-\tau)s^2+\frac{p^4-r^4}{4}\right].}
\end{equation}
Obviously, $R$ is a first integral of the induced system, and the
equation
\begin{equation}\label{s7n11}
\displaystyle{s^4-(p^2-\tau)s^2+\frac{p^4-r^4}{4}=0}
\end{equation}
defines the set of the integral constants $s,\tau$ such that on the
corresponding integral manifolds the 2-form induced on
$\mathfrak{O}$ by the symplectic structure $\boldsymbol \lambda$
degenerates. As the functions (\ref{s2n11}), (\ref{s2n12}) have an
algebraic structure, the set (\ref{s7n11}) has codimension 1 in
$\mathfrak{O}$. Thus, $\boldsymbol \lambda | _{\mathfrak{O}}$ is
almost everywhere non-degenerate.

To locate the set (\ref{s7n11}) on the surface (\ref{s2n14}) in
${\bf R}^3(h,k,g)$ notice that, in virtue of (\ref{s2n14}),
\begin{equation}\label{s7n12} \notag
\frac{\partial}{\partial s}(h,k,g) \times \frac{\partial}{\partial
\tau}(h,k,g)
=\bigl[\displaystyle{s^4-(p^2-\tau)s^2+\frac{p^4-r^4}{4}}
\bigr](\frac{\tau-p^2}{2 s^4},\frac{1}{2 s^3},\frac{1}{s^4}).
\end{equation}
Therefore, the equation (\ref{s7n11}) defines cuspidal edges of the
surface (\ref{s2n14}).

\section{Bifurcation diagram and the existence of motions}
Introduce the integral map $J$ of the dynamical system on
$\mathfrak{O}$
\begin{equation}\label{s5n1}
J(\zeta ) = (S(\zeta ),{\rm T}(\zeta )) \in {\bf{R}}^2, \quad \zeta
\in \mathfrak{O}.
\end{equation}
The bifurcation diagram $\Sigma(J)$ of this map is, by definition,
the set of pairs $(s, \tau)$ over which $J$ is not locally trivial.
In our case all integral manifolds (\ref{s4n1}) are compact.
Therefore, $\Sigma(J)$ is the set of critical values of $J$.

Define {\it the admissible region} as the image of the map
(\ref{s5n1}), i.e., the set of values $(s,\tau)$ such that the
integral manifold (\ref{s4n1}) is not empty. Obviously, $\partial
J(\mathfrak{O}) \subset \Sigma(J)$.

Due to the above results the existence of a trajectory on
$\mathfrak{O}$ with given $s, \tau$ is equivalent to the existence
of a point on the $(x,\xi)$-plane for which the values
(\ref{s4n26})--(\ref{s4n28}), (\ref{s4n35}), (\ref{s4n35}) satisfy
(\ref{s4n7}). It easily follows from (\ref{s4n25}), (\ref{s4n28}),
(\ref{s4n8}) that the conditions (\ref{s4n7}), in turn, are
equivalent to the system of inequalities
\begin{eqnarray}
& & \Phi_1(x, \xi) \geqslant 0, \quad \Phi_2(x, \xi) \leqslant 0, \label{s5n2} \\
& & \Psi_1(x, \xi) \geqslant 0, \quad \Psi_2(x, \xi) \leqslant 0
\label{s5n3}
\end{eqnarray}
considered in the quadrant $x\geqslant 0$, $ s\,\xi \geqslant 0$.

\begin{theorem}
The bifurcation diagram $\Sigma(J)$ consists of the following
subsets of the $(s,\tau )$-plane:
$$
\begin{array}{l}
\;1^ \circ)\; \tau  = (a + b)^2 ,\;s \in [ - a,0) \cup [b, + \infty
);\\
\;2^ \circ)\; \tau  = (a - b)^2 ,\;s \in [ - a, - b] \cup (0, +
\infty );\\
\;3^ \circ)\; s =  - a,\;\tau  \geqslant (a - b)^2 ;\\
\;4^ \circ)\; s =  - b,\;\tau  \geqslant (a - b)^2 ;\\
\;5^ \circ)\; s = b,\;\tau  \leqslant (a + b)^2 ;\\
\;6^ \circ)\; s = a,\;\tau  \leqslant (a + b)^2 ;\\
\;7^ \circ)\; \tau  = 0,\;s \in (0, + \infty );\\
\;8^ \circ)\; \tau  = (\sqrt {a^2  - s^2} + \sqrt{b^2 - s^2 })^2 ,\;s \in [ - b,0);\\
\;9^ \circ)\; \tau  = (\sqrt {a^2  - s^2} - \sqrt{b^2 - s^2 })^2 ,\;s \in (0,b];\\
10^ \circ)\; \tau  = - (\sqrt {s^2  - a^2}-\sqrt{s^2 - b^2})^2 ,\;s
\in [a, + \infty ).
\end{array}
$$
\end{theorem}

\begin{theorem}
The solutions of the system {\rm{(\ref{s1n1})}} under the conditions
{\rm{(\ref{s1n12})}} exist iff the constants of the first integrals
{\rm{(\ref{s2n11})}}, {\rm{(\ref{s2n12})}} satisfy one of the
following
$$
\begin{array}{l}
1^ \circ)\; -a \leqslant s \leqslant -b, \; \tau  \geqslant (a -
b)^2;\\
2^ \circ)\; -b \leqslant s < 0,\; \tau  \geqslant (\sqrt{\mathstrut
a^2 - s^2}+\sqrt{\mathstrut b^2-s^2})^2;\\
3^ \circ)\;\; 0 < s \leqslant b, \;\tau  \leqslant (\sqrt{\mathstrut
a^2 - s^2}- \sqrt{\mathstrut b^2-s^2})^2;\\
4^ \circ)\;\; b \leqslant s \leqslant a, \;\tau  \leqslant (a +
b)^2;\\
5^ \circ)\;\;  s \geqslant a, \;-(\sqrt{\mathstrut s^2 - b^2}-
 \sqrt{\mathstrut s^2-a^2})^2 \leqslant \tau \leqslant (a+b)^2.
\end{array}
$$
\end{theorem}

The complete proof of these statements is purely technical (see
\cite{Kh35}) and contains the scrupulous analysis of the regions on
the $(x,\xi)$-plane defined by (\ref{s5n2}), (\ref{s5n3}) and the
cases of their structural transformations. Another approach is
suggested in \cite{KhShv} for the pair $(S,H)$. It is based on the
classification of the trajectories in $\mathfrak{O}$ satisfying the
condition ${\rm {rank}} \, (H \times K \times G)<2$.

\begin{figure}[ht]\label{fig1}
\centering
\includegraphics[width=8cm,keepaspectratio]{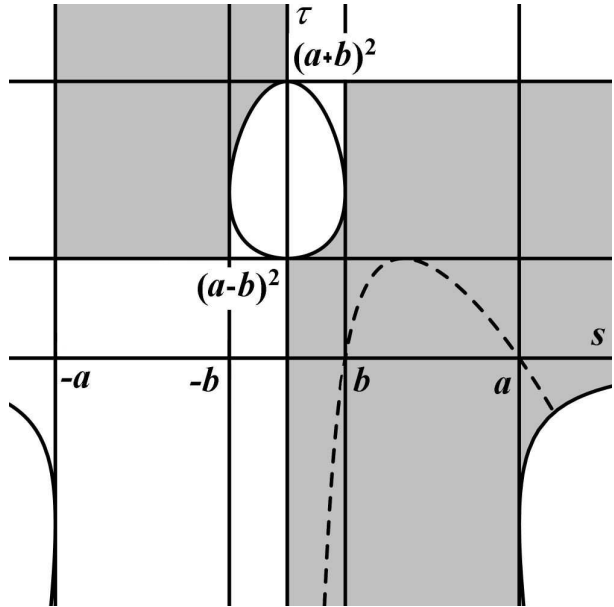}
\caption{The admissible region in the $(s,\tau)$-plane.}
\end{figure}

The admissible region is shaded in Fig.1. The dense lines and curves
represent the equations of the bifurcation diagram, the dashed curve
illustrates the equation (\ref{s7n11}), i.e., the first integrals
constants such that the symplectic structure is degenerate on the
corresponding integral manifolds.

\section{Separation of variables}
In the sequel we suppose that $\tau \sigma \ne 0$. In fact, if
$\tau=0$, then from (\ref{s2n14}) we obtain the relation
${(2g-p^2h)^2=r^4 k}$ characteristic for the critical manifold
$\mathfrak{N}$ \cite{KhOne}. The equations of motion on
$\mathfrak{N}$ were explicitly integrated in \cite{KhSavUMB}.
If $\sigma=0$, then the equations (\ref{s2n14}) yield one of the
relations (\ref{s1n11}). This case corresponds to the set of points
(\ref{s1n9}). At these points the manifold $\mathfrak{O}$ fails to
be smooth. The corresponding trajectories are the pendulum motions
(\ref{s1n10}).

Considering the second equation (\ref{s4n21}) denote $ \mu =
|r^2x_1-\tau y_1|=|r^2 x_2-\tau y_2|$. Then
\begin{equation}\label{s6n1}
\mu^2=\tau \xi^2+\sigma x^2-\tau\sigma.
\end{equation}
This equation defines the second-order surface $\mathcal{M}$ in
three-dimensional space ${\bf R}^3(x,\xi ,\mu )$. Each trajectory in
$\mathfrak{O}$ is in a natural way represented by a curve on
$\mathcal{M}$.

\begin{theorem}\label{th5}
Supposing $\tau \sigma \ne 0$, introduce the variables
\begin{equation}\label{s6n2}
\displaystyle{U=\frac{\tau \xi + x \mu}{\sqrt{\sigma}(\tau -
x^2)},\quad V=\frac{\tau \xi - x \mu}{\sqrt{\sigma}(\tau - x^2)}}.
\end{equation}
Then the equations of motion on $\mathfrak{O}$ separate
\begin{equation}\label{s6n3}
\displaystyle{\frac{dU}{\sqrt{Q(U)}} - \frac{dV}{\sqrt{Q(V)}} = 0,
\quad \frac{U dU}{\sqrt{Q(U)}} - \frac{V dV}{\sqrt{Q(V)}} = \frac
{dt}{\sqrt{\mathstrut 2s \tau \sigma}}}.
\end{equation}
Here
\begin{equation}\label{s6n4}
Q(w)=(w^2-1)({\sigma} w^2-{4s^2\chi^2})[(\sqrt{\sigma} w +
\tau)^2-r^4].
\end{equation}
\end{theorem}

\begin{proof}
Consider the local coordinates $u,v$ on the surface $\mathcal{M}$:
\begin{equation}\label{s6n5}
\xi  = \sqrt \sigma  \, \frac{uv + 1}{u + v},\quad x = \sqrt \tau \,
\frac{u - v}{u + v}, \quad \mu = \sqrt {\tau \sigma} \, \frac{uv -
1}{u + v}.
\end{equation}
In addition to (\ref{s4n19}) note the following two identities for
the constants introduced above,
\begin{equation}\label{s6n6}
\sigma  + 2\tau (p^2  \pm r^2 ) = (\tau  \pm r^2 )^2 .
\end{equation}
The polynomials (\ref{s4n15}), (\ref{s4n24}), (\ref{s4n36}) become
$$
\begin{array}{ll}
\displaystyle{\Phi_1  = \kappa\,\varphi _1 (u)\,\varphi _1 (v)}, &
\displaystyle{\Phi_2  = \kappa\,\varphi _2 (u)\,\varphi _2 (v),}
\\[2mm]
\displaystyle{\Psi_1  = \kappa\,\psi_1 (u)\, \psi_2 (v)}, &
\displaystyle{\Psi_2  = \kappa\,\psi_2(u) \,\psi_1(v),}
\\[2mm]
\displaystyle{\Theta_1  = \kappa\,\theta_2(u)\, \theta_1(v)}, &
\displaystyle{\Theta_2  = \kappa\,\theta_1(u) \,\theta_2(v),}
\end{array}
$$
where $\kappa=1/(u + v)^2$ and
$$
\begin{array}{ll}
\varphi _1 (w) = \sqrt \sigma  (1 + w^2 ) + 2(\tau  + r^2 )w, &
\varphi _2 (w) = \sqrt \sigma  (1 + w^2 ) + 2(\tau  - r^2 )w,
\\[2mm]
\psi _1 (w) = 2 s [(\chi+ \sqrt{\tau})w^2 - (\chi- \sqrt{\tau})], &
\psi _2 (w) = 2 s [(\chi- \sqrt{\tau})w^2 - (\chi+ \sqrt{\tau})], \\[2mm]
\theta_1 (w) = \sqrt \sigma  (1 - w^2 ) + 4 s \sqrt{\tau} w, &
\theta_2 (w) = \sqrt \sigma  (1 - w^2 ) - 4 s \sqrt{\tau} w.
\end{array}
$$
Then from (\ref{s4n26})--(\ref{s4n28}) we find the expressions for
the configuration variables
\begin{eqnarray}
& & \begin{array}{l} \displaystyle{ x_1  = \frac  {2s \tau}{r^2}
\left[ { \frac { \sqrt {\varphi _1 (u)\varphi _2 (v)} + \sqrt
{\varphi _2(u)\varphi _1 (v)} } {\sqrt {\theta _1 (u)\theta _1 (v)}
- \sqrt {\theta _2 (u)\theta _2 (v)}} } \, \right] ^2 , }  \\
\displaystyle{ x_2  = \frac {2s\tau } {r^2 } \left[{ \frac
{\sqrt{\varphi _1 (u)\varphi _2 (v)}  - \sqrt {\varphi _2 (u)\varphi
_1 (v)} } {\sqrt {\theta _1 (u)\theta _1 (v)}  + \sqrt {\theta _2
(u) \theta _2 (v)} }} \, \right]^2 ,}
\end{array} \label{s6n7} \\
& & \begin{array}{l} \displaystyle{ y_1  = 2s \frac {\left[ {{\sqrt
{\varphi _1 (u) \varphi _2 (v)}  + \sqrt {\varphi _2 (u)\varphi _1
(v)} } } \,\right]^2 - 4\sigma (uv - 1)^2 } {\left[ {\sqrt {\theta
_1 (u)\theta _1 (v)}  - \sqrt {\theta _2 (u)\theta _2 (v)} }
\,\right]^2
}\,,}  \\
\displaystyle { y_2  = 2s \frac {\left[ {\sqrt {\varphi _1
(u)\varphi _2 (v)}  - \sqrt {\varphi _2 (u)\varphi _1 (v)} }
\,\right]^2  - 4\sigma (uv - 1)^2 } {\left[ {\sqrt {\theta _1
(u)\theta _1 (v)}  + \sqrt {\theta _2 (u)\theta _2 (v)} }
\,\right]^2 }\,,}
\end{array} \label{s6n8}\\
& & \begin{array}{c} \displaystyle {z_1  = \frac{1}{2r(u + v)}
\left[ {\sqrt {\varphi _1 (u)\varphi _1 (v)}  + \sqrt {\varphi _2
(u)\varphi _2 (v)} } \,\right]\,,} \\
\displaystyle {z_2  = \frac{1}{2r(u + v)} \left[ {\sqrt {\varphi _1
(u)\varphi _1 (v)}  - \sqrt {\varphi _2 (u)\varphi _2 (v)} }
\,\right]\,.}
\end{array}\label{s6n9}
\end{eqnarray}
Hence, in particular,
\begin{equation}\label{s6n10}
\begin{array}{l}
\displaystyle{\sqrt{x_1} = \frac{\sqrt{\mathstrut 2 s \, \tau}}{r}
\frac {\sqrt{\varphi_1 (u)\varphi_2 (v)}+\sqrt{\varphi_2
(u)\varphi_1 (v)}} {\sqrt{\theta_1 (u)\theta_1 (v)}-\sqrt{\theta_2
(u)\theta_2 (v)}}}, \\[4mm] \displaystyle{\sqrt{x_2} =
\frac{\sqrt{\mathstrut 2 s \, \tau}}{r} \frac {\sqrt{\varphi_1
(u)\varphi_2 (v)}-\sqrt{\varphi_2 (u)\varphi_1 (v)}} {\sqrt{\theta_1
(u)\theta_1 (v)}+\sqrt{\theta_2 (u)\theta_2 (v)}}}.
\end{array}
\end{equation}
Here the arbitrary choice of sign is provided by the algebraic value
$\sqrt{2s\tau}$. From (\ref{s4n35}) we have
\begin{equation}\label{s6n11}
\begin{array}{l}
\displaystyle{\sqrt{x_2} w_1  = \frac{i}{4 s} \frac
{\sqrt{\theta_2(u)\theta_1(v)} +
\sqrt{\theta_1(u)\theta_2(v)}}{u+v}}, \\[4mm]
\displaystyle{\sqrt{x_1} w_2 = \frac{i}{4s} \frac
{\sqrt{\theta_2(u)\theta_1(v)} - \sqrt{\theta_1(u) \theta_2(v)} }
{u+v} }.
\end{array}
\end{equation}
Substitute (\ref{s6n10}) into (\ref{s6n11}) to obtain the
expressions for the variables defining the equatorial components of
the angular velocity
\begin{equation}\label{s6n12}
\begin{array}{l}
\displaystyle{w_1  = \frac {i r}{4s\sqrt {2s\tau}} {{\left[ {\sqrt
{\theta _2 (u)\theta _1 (v)}  + \sqrt {\theta _1 (u)\theta _2 (v)} }
 \right]\left[ {\sqrt {\theta _1 (u)\theta _1 (v)}  +
 \sqrt {\theta _2 (u)\theta _2 (v)} } \right]}
 \over {(u + v)
 \left[ {\sqrt {\varphi _1 (u)\varphi _2 (v)}  - \sqrt {\varphi _2 (u)\varphi _1 (v)} }
 \right]}},} \\
\displaystyle{w_2  = \frac {i r}{4s\sqrt {2s\tau}}{{\left[ {\sqrt
{\theta _2 (u)\theta _1 (v)}  - \sqrt {\theta _1 (u)\theta _2 (v)} }
\right]\left[ {\sqrt {\theta _1 (u)\theta _1 (v)}  - \sqrt {\theta
_2 (u)\theta _2 (v)} } \right]} \over {(u + v)\left[ {\sqrt {\varphi
_1 (u)\varphi _2 (v)}  + \sqrt {\varphi _2 (u)\varphi _1 (v)} }
\right]}}.}
\end{array}
\end{equation}
The axial component is found from (\ref{s4n35}),
\begin{equation}\label{s6n13}
w_3  = \frac{i}{2\sqrt {2s\tau \sigma}} \frac { {\sqrt {\theta _1
(u)\theta _2 (u)\varphi _1 (v)\varphi _2 (v)} - \sqrt {\varphi _1
(u)\varphi _2 (u)\theta _1 (v)\theta _2 (v)} } }{(u + v)(uv - 1)}.
\end{equation}
Thus we have expressed all phase variables in terms of $u,v$.

To obtain the differential equations for $u,v$ consider the
following variables
\begin{equation}\label{s6n14}
s_1  = \frac{x^2  + z^2  + r^2}{2x},\quad s_2  = \frac{x^2  + z^2 -
r^2 }{2x}.
\end{equation}
The derivatives, in virtue of the system (\ref{s1n8}), are
\begin{equation}\label{s6n15}
\displaystyle{s_1'= \frac{r^2}{4x^3}(z_1+z_2)(x_1 w_2 - x_2
w_1),\quad s_2'= \frac{r^2}{4x^3}(z_1-z_2)(x_1 w_2 + x_2 w_1).}
\end{equation}
On the other hand, from (\ref{s6n14}), (\ref{s6n5}) we have
$$
\displaystyle{s_1= \frac{\sqrt{\sigma}(u v+1)+(\tau+r^2)(u+v)}{2
\sqrt{\tau}(u-v)},} \quad \displaystyle{s_2= \frac{\sqrt{\sigma}(u
v+1)+(\tau-r^2)(u+v)}{2 \sqrt{\tau}(u-v)},}
$$
whence
$$
\begin{array}{ll}
\displaystyle{\frac{\partial s_1}{\partial u}=-\frac {\varphi_1
(v)}{2 \sqrt{\tau}(u-v)^2},} & \displaystyle{\frac{\partial
s_1}{\partial v}= \frac
{\varphi_1 (u)}{2 \sqrt{\tau}(u-v)^2},}\\
\displaystyle{\frac{\partial s_2}{\partial u}=-\frac {\varphi_2
(v)}{2 \sqrt{\tau}(u-v)^2},} & \displaystyle{\frac{\partial
s_2}{\partial v}= \frac {\varphi_2 (u)}{2 \sqrt{\tau}(u-v)^2}.}
\end{array}
$$
Therefore,
\begin{equation}\label{s6n16}
\begin{array}{l}
\displaystyle{\frac{du}{dt}=\frac{2
\sqrt{\tau}(u-v)^2}{\varphi_1(u)\varphi_2(v)-\varphi_2(u)\varphi_1(v)}
[\varphi_2(u)\frac{ds_1}{dt}-\varphi_1(u)
\frac{ds_2}{dt}],}\\[4mm]
\displaystyle{\frac{dv}{dt}=\frac{2
\sqrt{\tau}(u-v)^2}{\varphi_1(u)\varphi_2(v)-\varphi_2(u)\varphi_1(v)}
[\varphi_2(v)\frac{ds_1}{dt}-\varphi_1(v) \frac{ds_2}{dt}].}
\end{array}
\end{equation}
Substitute the values (\ref{s6n9})--(\ref{s6n11}) into (\ref{s6n15})
and the resulting expressions into (\ref{s6n16}). We obtain
\begin{equation}\label{s6n17}
\begin{array}{l}
f(u,v)\,\displaystyle{\frac{du}{dt}=\frac{\sqrt {\mathstrut
\varphi_1(u)\varphi_2(u) \, \theta_1(u) \, \theta_2(u)}}{2 u
\sqrt{\mathstrut 2 s\,\tau \sigma}}}, \quad
f(u,v)\,\displaystyle{\frac{dv}{dt}=\frac{\sqrt{\mathstrut
\varphi_1(v)\varphi_2(v)\,\theta_1(v)\,\theta_2(v)}}{2 v
\sqrt{\mathstrut 2 s\,\tau \sigma}}},
\end{array}
\end{equation}
where
$$
\displaystyle{f(u,v)=\frac{(u-v)(1-u v)}{u
v}=\Bigl(v+\frac{1}{v}\Bigr)-\Bigl(u+\frac{1}{u}\Bigr)}.
$$

The variables (\ref{s6n2}) in terms of $u,v$ have the form
\begin{equation}\label{s6n18}
U=\frac{1}{2}\Bigl(u+\frac{1}{u}\Bigr), \quad V=\frac{1}{2}\Bigl(v+\frac{1}{v}\Bigr).
\end{equation}
To be definite choose $u=U-\sqrt{\mathstrut U^2-1}$,
$v=V-\sqrt{\mathstrut V^2-1}$. Then the equations (\ref{s6n17})
yield
\begin{equation}\label{s6n19}
(U-V)\frac{dU}{dt}=\frac{1}{\sqrt{2s \, \tau
\sigma}}\sqrt{Q(U)},\quad (U-V)\frac{dV}{dt}=\frac{1}{\sqrt{2s \,
\tau \sigma}}\sqrt{Q(V)}
\end{equation}
with the polynomial (\ref{s6n4}) of degree 6. This system is
obviously equivalent to the system (\ref{s6n3}); the latter have the
standard form for hyperelliptic quadratures.
\end{proof}

To reveal the connection of Theorem~\ref{th5} with the bifurcation
diagram suppose that the polynomial (\ref{s6n4}) has a multiple
root. The resultant of $Q(w)$ and $Q'(w)$ is
\begin{equation}\label{s6n20}
2^{32} a^4 b^4 (a^2-b^2)^2 s^{10} \tau^{12} \sigma^8 \chi^2
(a^2-s^2)^2(b^2-s^2)^2.
\end{equation}
According to (\ref{s2n14}), $s \ne 0$ on $\mathfrak{O}$. The
equation $\sigma=0$ gives $\tau=(a \pm b)^2$. The case $\chi =0$
provides the values $ \tau =(\sqrt{\mathstrut a^2-s^2} \pm
\sqrt{\mathstrut b^2-s^2})^2$. Thus, the bifurcation diagram of
${J=S \times \mathrm{T}}$ belongs to the discriminant set of the
polynomial (\ref{s6n4}). This fact is typical for systems with
separating variables.

Note that all roots of the polynomial (\ref{s6n4}) are explicitly
expressed in terms of the integral constants along with the roots of
the polynomial
\begin{equation}\label{s6n21}
\varphi_1(w)\varphi_2(w) \, \theta_1(w) \, \theta_2(w)
\end{equation}
defining the solutions of the system (\ref{s6n17}). The variables
introduced in this section are real only in the case $\tau
>0$ and $\sigma >0$. For other combinations of signs the definition (\ref{s6n5})
of the local coordinates $u,v$ needs obvious modifications. Then the
intervals of oscillations for the variables $U,V$ and $u,v$ are
easily determined for any given values of $s,\tau$ or, more exactly,
for each connected component of ${\bf R}^2\backslash \Sigma(J)$. The
obtained differential equations and the explicit formulae for the
phase variables in terms of $u,v$ provide the possibility to
investigate the phase topology of the case considered and to
calculate analytically the numerical invariants of the corresponding
Liouville foliation.


\begin{thebibliography}{99}
\bibitem{KhRCD} M.\,P.\,Kharlamov.
Bifurcation diagrams of the Kowalevski top in two constant fields.
{\it Regul. Chaotic Dyn.}, {\bf 10} (2005), 381--398.

\bibitem{Bogo} O.\,I.\,Bogoyavlensky. Euler equations on
finite-dimension Lie algebras arising in physical problems. {\it
Commun. Math. Phys.}, {\bf 95} (1984), 307--315.

\bibitem{ReySem} A.\,G.\,Reyman, M.\,A.\,Semenov-Tian-Shansky. Lax representation
with a spectral parameter for the Kowalewski top and its
generalizations. {\it Lett. Math. Phys.}, {\bf 14} (1987), 55--61.

\bibitem{KhMTT} M.\,P.\,Kharlamov. Critical set and bifurcation diagram
of the problem of motion of the Kowalevski top in double field. {\it
Mech. Tverd. Tela}, \No\,34, 2004, 47--58. (In Russian)

\bibitem{Kowa}
S.\,Kowalevski. Sur le probl\`{e}me de la rotation d'un corps solide
autour d'un point fixe. {\it Acta Math.}, {\bf 12} (1889), 177--232.

\bibitem{Appel} G.\,G.\,Appelrot. Non-completely symmetric heavy
gyroscopes. In: {\it Motion of a rigid body about a fixed point}.
Collection of papers in memory of S.V.Kovalevskay. Acad. Sci. USSR,
Moscow-Leningrad, 1940, 61--156. (In Russian)

\bibitem{ZotRCD} D.\,B.\,Zotev. Fomenko-Zieschang invariant in the Bogoyavlenskyi
case. {\it Regul. Chaotic Dyn.}, {\bf 5} (2000), 437--458.

\bibitem{KhSavUMB} M.\,P.\,Kharlamov, A.\,Y.\,Savushkin.
Separation of variables and integral manifolds in one partial
problem of motion of the generalized Kowalevski top. {\it Ukr. Math.
Bull.}, {\bf 1} (2004), 548--565.

\bibitem{KhOne}
M.\,P.\,Kharlamov. One class of solutions with two invariant
relations in the problem of motion of the Kowalevsky top in double
constant field. {\it Mekh. tverd. tela}, \No\,32, 2002, 32--38. (In
Russian)

\bibitem{Kh35}
M.\,P.\,Kharlamov. Bifurcation diagram of the generalized 4th
Appelrot class. {\it Mekh. tverd. tela}, \No\,35, 2005, 38--48. (In
Russian)

\bibitem{KhShv} M.\,P.\,Kharlamov, E.\,G.\,Shvedov.
On the existence of motions in the generalized 4th Appelrot class.
{\it Regul. Chaotic Dyn.}, {\bf 11} (2006), 337--342.

\end{thebibliography}
\end{document}